# Correlating grain boundary character and composition in 3-dimensions using 4D-scanning precession electron diffraction and atom probe tomography

Saurabh M. Das[1*+], Patrick Harrison[2+], Srikakulapu Kiranbabu[1], Xuyang Zhou[1], Wolfgang Ludwig[3,4*], Edgar F. Rauch[2], Michael Herbig[1], Christian H. Liebscher[1,5,6*]

[1]Max-Planck-Institut for Sustainable Materials (Max-Planck-Institut für Eisenforschung), Max-Planck-Straβe 1, 40237 Düsseldorf, Germany
[2]SIMAP Laboratory, CNRS-Grenoble INP, BP 46 101 rue de la Physique, 38402 Saint Martin d'Hères, France
[3]ESRF–The European Synchrotron, 71 Av. des Martyrs, 38000 Grenoble, France
[4]MATEIS, INSA Lyon, UMR 5510 CNRS, 25 av Jean Capelle, 69621 Villeurbanne, France
[5]Research Center Future Energy Materials and Systems, Ruhr Univeristy Bochum, Universitätsstr. 150, 44801 Bochum, Germany
[6]Faculty of Physics and Astronomy, Ruhr Univeristy Bochum, Universitätsstr. 150, 44801 Bochum, Germany

*Corresponding authors,

email: 100rabh2992@gmail.com (Saurabh Mohan Das), wolfgang.ludwig@esrf.fr (Wolfgang Ludwig),

christian.liebscher@rub.de (Christian H. Liebscher)

[+]These authors contributed to the work equally.

**Abstract**

Grain boundaries are dominant imperfections in nanocrystalline materials that form a complex 3-dimensional (3D) network. Solute segregation to grain boundaries is strongly coupled to the grain boundary character, which governs the stability and macroscopic properties of nanostructured materials. Here, we develop a 3-dimensional transmission electron microscopy and atom probe tomography correlation framework to retrieve the grain boundary character and composition at the highest spatial resolution and chemical sensitivity by correlating four-dimensional scanning precession electron diffraction tomography (4D-SPED) and atom probe tomography (APT) on the same sample. We obtain the 3D grain boundary habit plane network and explore the preferential segregation of Cu and Si in a nanocrystalline Ni-W alloy. The correlation of structural and compositional information reveals that Cu segregates predominantly along high angle grain boundaries and incoherent twin boundaries, whereas Si segregation to low angle and incommensurate grain boundaries is observed. The novel full 3D correlative approach employed in this work opens up new possibilities to explore the 3D crystallographic and compositional nature of nanomaterials. This lays the foundation for both probing the true 3D structure-chemistry at the sub-nanometer scale and, consequentially, tailoring the macroscopic properties of advanced nanomaterials.

**Main**

The structure and composition of interfaces such as phase boundaries, grain boundaries (GBs), and twin boundaries play a vital role in controlling the macroscopic properties of polycrystalline materials. Moreover, the behavior of materials such as thermal stability, fracture toughness, electrochemical properties, hydrogen embrittlement, and electrical conductivity can be tailored through the design of interface structure and composition[1–3]. Generally, the structure of any interface that separates differently oriented crystals, termed grain boundary, can be described by five independent crystallographic parameters (macroscopic degrees of freedom, DOF). Three of them describe the mutual misorientation of the adjoining crystals and two define the inclination of the interface (interface habit plane). The enrichment of solute elements at these interfaces, termed segregation, adds another level of complexity to the determination of interfacial, and hence material properties. Grain boundaries typically span a complex three-dimensional network within polycrystalline materials with locally varying interface parameters confined to within several nanometers. As a result, due to the locality and scale of these interfaces, the development of novel characterization techniques are required to retrieve their structural and compositional information in three dimensions and at the highest possible spatial and compositional resolution.

Structural information on crystal orientations and interfaces is typically resolved down to atomic resolution by Scanning and Transmission Electron Microscopy (SEM and TEM). Electron Backscatter Diffraction (EBSD) in the SEM and Scanning Precession Electron Diffraction (SPED) in the TEM are widely used to reveal crystal misorientations and geometric parameters of grain boundaries from single projections[4,5]. Atomic resolution scanning transmission electron microscopy is capable of resolving the atomic nature of grain boundaries, however, mostly limited to special grain boundaries with a common tilt or twist axis[6,7]. Three-dimensional (3D) techniques, such as 3D X-ray Diffraction Microscopy or serial sectioning EBSD, have been established to determine the complete crystallography of polycrystalline materials and the grain boundaries therein with millimeter down to sub-micrometer resolution[8–11]. It has recently been demonstrated that the 3D atomic structure of GBs can be resolved by atomic electron tomography with sub-Angstrom resolution in a relatively small field of view of several tens of square nanometers[12]. Over the past decade, two TEM based orientation mapping techniques, dark-field conical scanning and SPED, have been coupled with tomography tilt series acquisition to obtain 3D crystal orientations with nanometer resolution[13–16]. While these techniques provide high resolution and 3D structural information, the associated local 3D distribution of elements at the interfaces remains elusive.

Spectroscopic methods in electron microscopy, such as Energy Dispersive X-ray (EDX) or Electron Energy Loss Spectroscopy (EELS), probe compositional information with high spatial resolution from 2D projections of a sample, however, are limited by their ability to detect low elemental concentrations or light elements[17]. Atom Probe Tomography (APT), on the other hand, can obtain 3D elemental distributions of materials with both high spatial resolution, down to the sub-nanometer level, and high chemical sensitivity, at the parts-per-million (ppm) level[18]. Microstructural features such as dislocations, stacking faults, and grain boundaries can only be indirectly revealed if solute elements are segregating or depleting at these material imperfections[19,20]. Crystallographic information can also be obtained by APT, however, in very limited scenarios and usually restricted to grain orientation analysis[21,22]. To determine both crystallographic and compositional information with highest possible resolution from the same sample location, techniques correlating electron microscopy with APT have been developed. Seminal correlative approaches were performed in the 1960s by including target preparation of grain boundary regions[23–25]. Refined sample preparation strategies have enabled a direct correlation of TEM and APT on the same sample[26–28]. These correlative approaches have provided novel insights into a wide variety of material science related phenomena, more specifically grain boundary segregation engineering (GBSE)[29–31]. Herbig et. al. have demonstrated that by combining scanning nanobeam diffraction mapping and APT, the local 2D grain boundary crystallography and 3D elemental distribution can be correlated in a nanocrystalline sample[29]. However, these correlative TEM/APT techniques can only fully characterize the 5-DOFs for planar interfaces parallel to the electron beam direction as exemplified for columnar nanocrystalline pearlitic steel or line segregation at faceted Si grain boundaries[29,32]. The vast majority of nanocrystalline materials exhibit more complex 3D grain shapes and boundary networks, which could not be accessed by correlative techniques, so far.

Direct correlation of 3D grain orientation mapping in the TEM and APT can mitigate the limitations of the individual techniques and enable a holistic characterization of complex nanomaterials in 3D with nanometer resolution. Here, we develop a full 3D correlative approach combining 4D-scanning precession electron diffraction tomography with atom probe tomography on the same sample. This technique enables to obtain 3D crystal shapes and orientations and a complete 5 parameter description of local grain boundary properties, as well as their associated 3D segregation levels in nanomaterials. We apply this method to nanocrystalline Ni-W alloys electrodeposited on Cu substrate due to their exceptional coarsening resistance and propensity to form equiaxed nano-grains. Our 3D correlative approach enables systematic analysis of the interrelation of local grain boundary crystallography and elemental segregation of Cu and Si with unprecedented resolution, thus paving the way towards integrated structure-composition characterization of nanomaterials.

**Results**

To link grain boundary character and segregation with highest possible spatial and chemical resolution in 3D, we correlate 4D-SPED tomography (4D-SPEDT) and 3D-APT on the same sample. The three main steps of the workflow are shown in Fig 1: 1) Acquisition of a 4D-SPEDT (see Figs. 1a and b) series for grain orientation and grain boundary reconstruction, 2) 3D-APT (Fig. 1c) to obtain compositional information from the same sample, and 3) data processing to combine the grain orientation and grain boundary crystallography from 4D-SPEDT with the elemental distribution from APT.

**Figure 1: The data acquisition framework for 3-dimensional structure-chemistry correlation.**

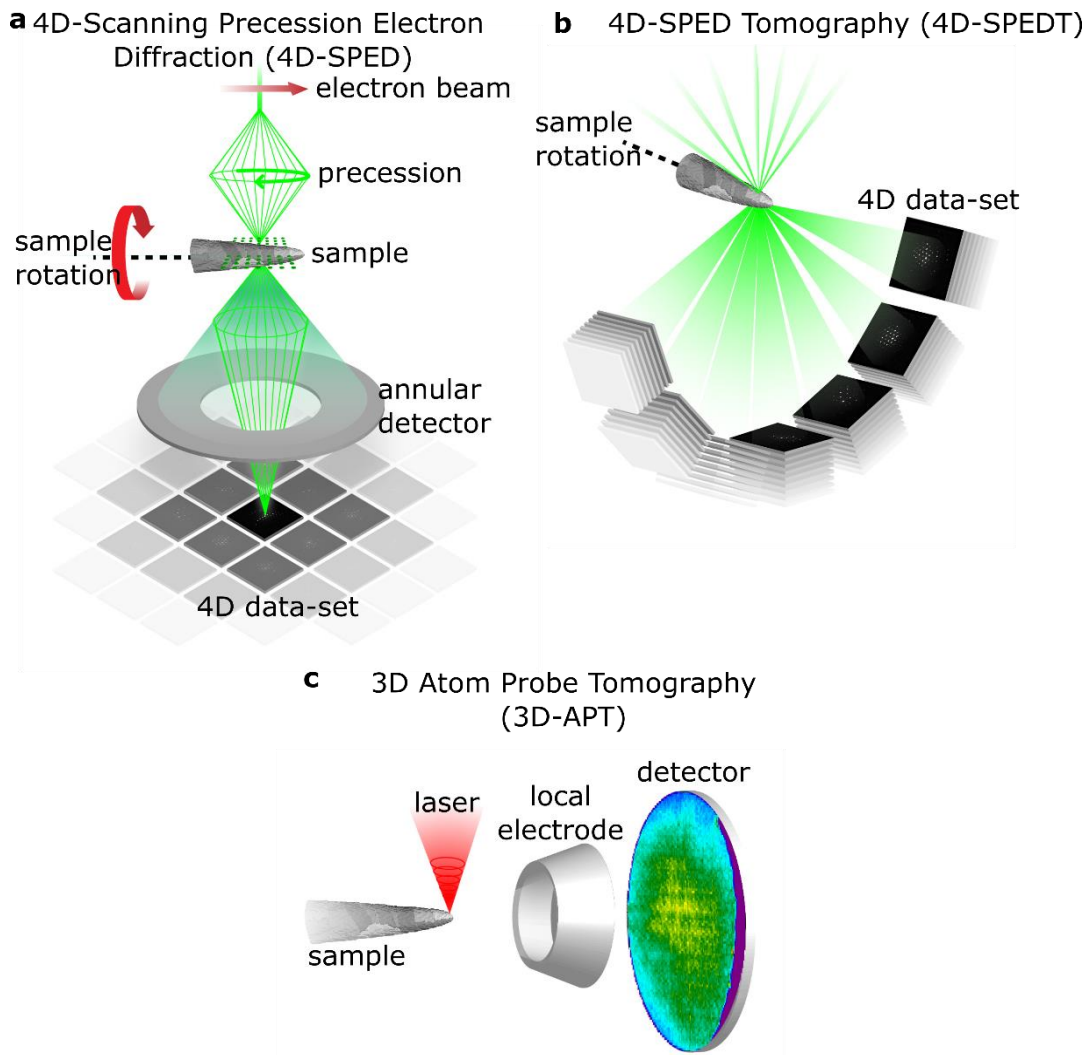

**Figure 1:** *Schematic illustration of the data acquisition protocol. (a) Four-dimensional scanning precession electron diffraction (4D-SPED) technique where a precessed nano-sized beam is scanned over a needle-shaped sample and electron diffraction patterns are recorded pixel by pixel using a pixelated detector. (b) The 4D-SPED dataset is acquired tilt by tilt in the range from -80° to +80° at a tilt step of 10° using an on-axis rotation tomographic holder. The sample is cleaned in a low-kV Argon ion shower before being loaded into the APT chamber. (c) The sample is field evaporated using electric and laser pulses and atoms are collected on a position sensitive detector.*

**Data Acquisition for 3D structure-composition correlation**

To determine the 3D crystallographic information (e.g. 5 DOFs of the grain boundaries), 4D-SPED was performed at each tilt angle throughout the tilt series. In this technique (Fig. 1a), a nano-sized incident beam is precessed by 1° whilst being scanned over a needle-shaped sample of a Ni-W nanocrystalline alloy to realize quasi-kinematical scattering conditions. The diffraction patterns are recorded at each probe position using a scintillator-coupled complementary metal–oxide–semiconductor (CMOS) camera. The resulting 4D dataset contains spatially resolved crystallographic information. A 4D dataset was recorded at each 10º tilt increment over a ±80° tilt range, as illustrated in Fig 1b. The same needle-shaped sample was subsequently measured by APT to obtain the 3D distribution of elements in the sample as illustrated in Fig 1c.

**Three-dimensional (3D) Crystal Orientation Mapping**

Automated crystal orientation mapping (ACOM) was used to obtain the crystal orientation on a pixel-by-pixel basis for each tilt dataset by comparing the experimental nanobeam diffraction patterns to a simulated template library[33]. Figure 2a illustrates the correlation between the optimal simulated diffraction template (open red circle overlay) and an experimental diffraction pattern acquired in the 60° tilt dataset. In a nanocrystalline sample, complex diffraction patterns may emerge at specific beam positions due to the possibility of overlapping grains along the beam direction. This is seen by the extra diffraction spots not matching the template in Figure 2a. A multi-indexing extension of ACOM has been employed to retrieve information from overlapping grains[34]. In this algorithm, the reflections corresponding to the best-matching template from the first indexation step are removed and the remaining diffraction spots are re-indexed, revealing multiple crystal orientations in the same diffraction pattern. This iterative process revealed the orientation of all overlapping grains. Supplementary Fig. S1 exemplifies this strategy for the 60° tilt dataset.

The projected images of individual grains were then obtained by using virtual dark field imaging[35,36], as shown in Fig 2b. Here, multiple virtual apertures are positioned around the experimentally obtained diffraction spots according to best-matched diffraction templates as shown in Fig 2a. Only the diffraction signal within these virtual apertures is integrated to reconstruct the corresponding projected image of a grain (see Fig. 2b)[16,37]. To account for the presence of twins in the sample, a twin orientation specific template scheme was used to automatically separate twin from matrix reflections, as illustrated in supplementary Figure S2. The separation of twin and matrix projection signal is a prerequisite for consistent tomographic shape reconstruction in materials exhibiting this type of orientation relationships[38]. As an example, an isolated grain containing twin domains was tracked for the entire tilt range as shown in Fig 2c.

Prior to 3D reconstruction of each grain, intensity normalizations accounting for the decay of the electron probe and for differences in diffraction conditions between different tilt settings were applied. The series of projected images for each grain was first aligned in the IMOD software using a global set of alignment parameters[39]. Subsequently, the grains were individually reconstructed in 3D using the Simultaneous Iterative Reconstruction Technique (SIRT) algorithm with 15 iterations combined with a non-negative minimum constraint to promote physical solutions. The 3D reconstruction of the grain shown in Fig. 2c is illustrated in Fig. 2d. Each grain was reconstructed independently using the global set of alignment parameters and located into a final common reconstruction volume. Hence, the reconstructed volume of all crystals in the dataset was colored according to their crystallographic orientation along z-axis as shown in Figure 2e (video in supplementary video 1) for two different view angles.

**Figure 2: Data processing protocol to retrieve through thickness orientation, grain shape through tilt series, and grain boundary habit plane normal distribution**

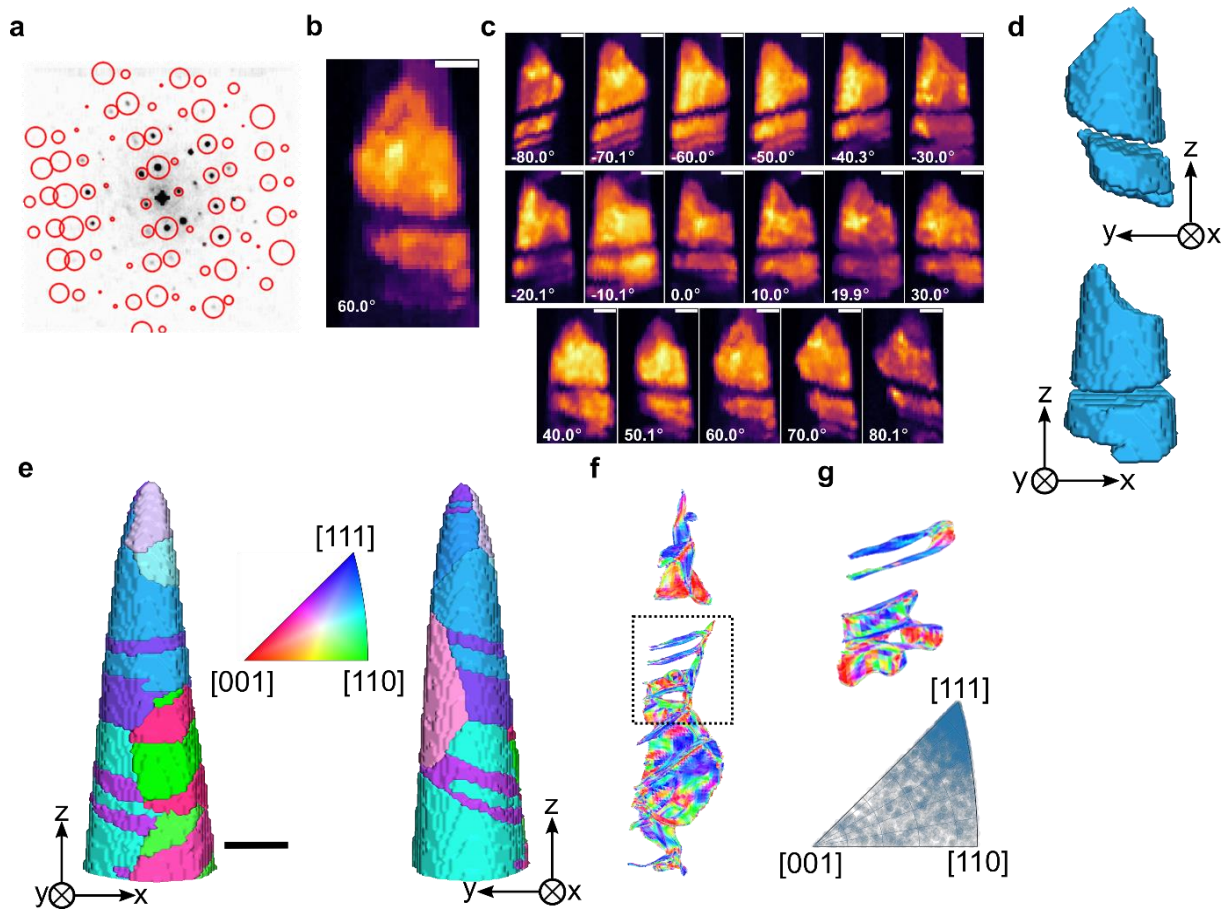

**Figure 2:** *(a) ACOM illustration demonstrating the correlation between the best-matched diffraction template (open red circle) and an experimental diffraction pattern extracted from 60° tilt dataset. (b) The orientation-specific virtual dark field image calculated using these virtual apertures (open red circle). (c) Orientation-specific virtual dark field images of a grain tracked through the entire tilt series. (d) The 3D reconstructed isosurface rendering of this grain in two different view angles. (e) The 3D reconstructed isosurface rendering of the Ni-W nanocrystalline alloy in two different view angles. Each grain is colored by its average orientation projected along the z-axis. (f) Grain boundary habit planes were reconstructed from the 3D reconstructed volume shown in (e) and segmented into triangular mesh. Mesh*

*faces are colored according to their crystallographic plane orientation in inverse pole figure color. (g) The grain boundaries from the highlighted (black dashed line) region in (f) and its grain boundary habit plane normal distribution plot. Scale bars are 50 nm.*

**Three-dimensional (3D) grain boundary mapping**

The reconstructed volume (256×74×74 voxel grid) shown in Fig. 2e consists of nano-sized Ni-W grains which are separated by general grain boundaries as well as coherent and incoherent twin boundaries. Since the crystallographic orientation of each grain is known, the grain misorientation and grain boundary plane normal can be obtained after conversion of the voxelated data into a 3D grain boundary mesh. Details on the procedure using a marching cubes algorithm[39] are described in the method section. The 3D grain boundary normal orientation plot is thus represented as a surface plot with two adjacent surfaces (see Figs. 2f and g). The local grain boundary normal direction is converted to the adjacent crystal reference frames and colored following the crystallographic inverse pole figure (IPF) color code as shown in Fig 2e. A zoomed in view of the grain boundary normal plot containing the twin boundaries (Figs. 2a-d) together with the corresponding grain boundary plane normal distribution (GBND) is shown in Fig 2g. The color of the two twin boundary surfaces suggests a <111> orientation of the surface, respectively, interface normal, corresponding to a coherent Σ3 twin boundary in a face centered cubic (fcc) crystal. This is confirmed when looking at the grain boundary plane normal distribution (GBND) shown in the bottom part of Fig 2g. Since this section of the reconstructed volume (highlighted in Fig. 2f) is populated with Σ3 annealing twin boundaries with a common {111} twin boundary plane, the GBND shows a preferred distribution of the interface planes of <111>.

**Correlative three-dimensional atom probe tomography**

The same sample was subsequently field-evaporated (electric and laser pulse) in the analysis chamber of an APT instrument, as detailed in the method section. To obtain an accurate 3D reconstruction, the reconstruction parameters need to be carefully calibrated. Gault et al. developed an approach to calibrate the parameters on lattice planes when multiple crystallographic poles are imaged within a single grain in the APT data[40]. However, the sample investigated in this study is a complex alloyed material system with a nanograined structure and obtaining crystallographic poles throughout the entire volume is challenging[41]. When selecting a specific ion evaporation sequence from the voltage history curve (marked by black dashed rectangular box in Fig. 3a), which corresponds to a grain in the 3D reconstructed volume shown in Fig. 3d, it is possible to reveal a single pole in the corresponding detector event histogram (see Fig. 3b). The reconstruction can be calibrated through choosing a cylindrical region (~2 nm diameter and ~10 nm depth) around the pole on the detector event histogram as indicated in Fig. 3c. This enables to resolve the {111} lattice planes in the $z$-spatial distribution map (SDM). We utilize the 4D-SPED data to confirm the orientation of the {111} lattice planes for this particular crystal. Hence, the optimized reconstruction parameters are applied to obtain the entire 3D atom map as shown in Fig 3d (video in supplementary video 2). It becomes apparent that Cu and Si atoms are predominantly present at the grain boundaries, which would otherwise be difficult to identify in APT, as illustrated in Fig. 3d.

**Figure 3: Atom probe tomography tip calibration, 3D chemistry, and crystallography**

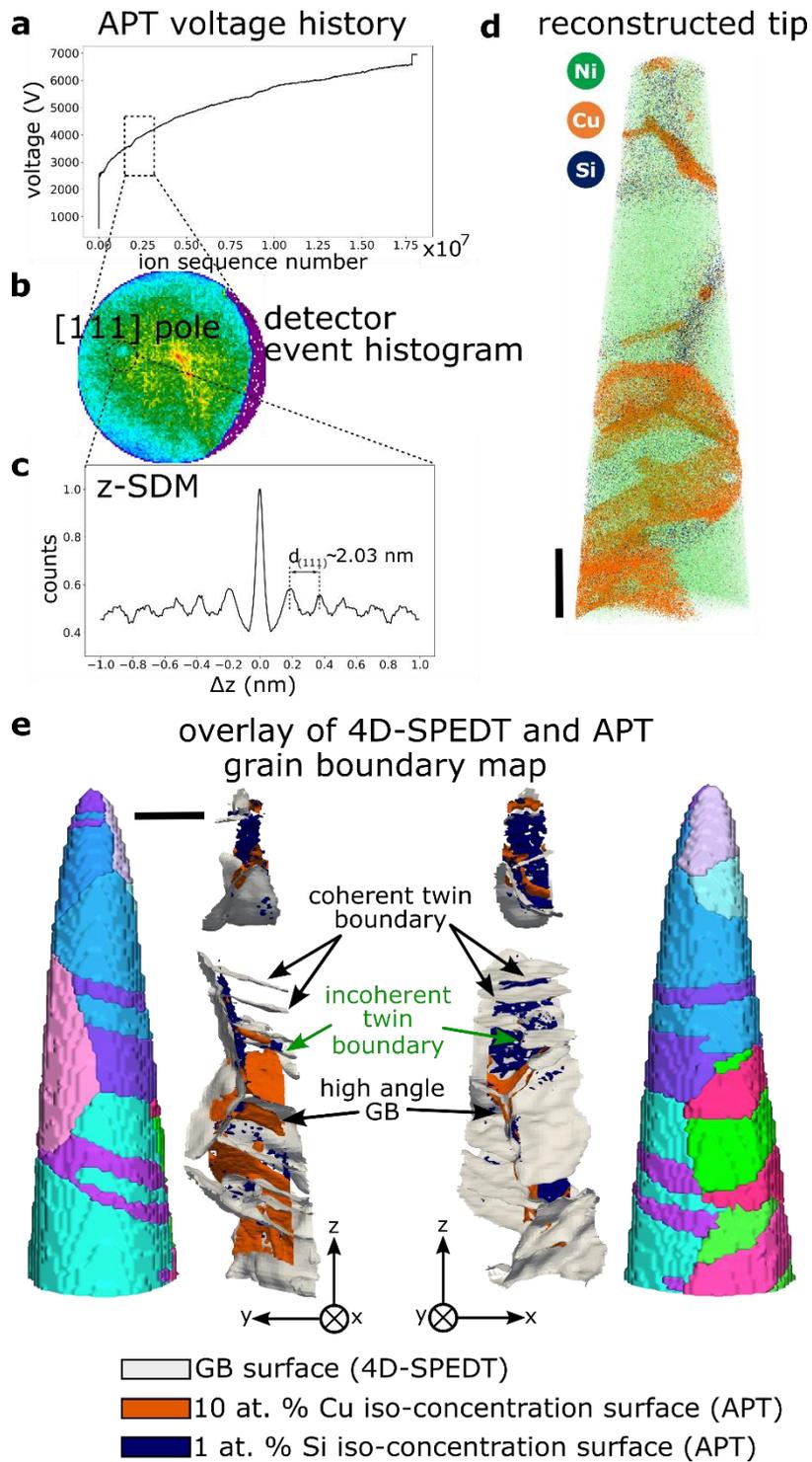

**Figure 3:** *(a) The voltage curve for the sample until fracture. The region highlighted by the black rectangle was used to generate the detector event histogram in (b). (b) The detector event histogram with multiple ion events revealing a crystallographic pole, here <111>, along the tip axis (z-axis). (c) Spatial distribution map of all ions in the selected sub-volume enabling to refine the reconstruction parameters to match the spacing of the lattice planes to {111} fcc of*

*Ni. (d) 3D atom map of the reconstructed volume displaying the distribution of Ni, Cu, and Si atoms. (e) Overlay of iso-concentration surfaces of Cu (10 at%, orange) and Si (1 at%, blue) obtained by APT on the grain boundary surface network (gray surface) reconstructed from 4D-SPED tomography (shown in Figure 2e) for two different view angles. The corresponding 3D crystal orientation map is shown for comparison. Scale bars are 50 nm.*

Since we now have the 3D crystal orientation and the corresponding 3D elemental distribution available from the same sample, it is possible to explore the 3D character of grain boundary segregation. To isolate grain boundary segregation in the APT data, iso-concentration surfaces of ~10 at% Cu (orange) and ~1 at% Si (blue) are extracted and overlaid on the reconstructed grain boundary network (gray surface) from the 4D-SPEDT data as shown in Fig. 3e. The entire grain reconstruction (in IPF color coding) from 4D-SPEDT is also shown alongside in the same orientation for comparison in Fig. 3e. This illustrates that overall, there is a good correlation between the 4D-SPEDT and APT datasets, especially considering that the APT reconstruction was refined using crystallographic information, but no further processing was applied. On the bottom side of the reconstruction away from the tip apex, one can observe a slightly larger deviation between the grain boundary surface reconstruction and the iso-concentration surfaces, which is related to limitations in the APT reconstruction.

**Linking grain boundary character and composition in 3D**

To demonstrate the direct 3D linking of the grain boundary character with solute segregation, we focus on two selected parts of the reconstruction shown in Fig. 3e. The grain reconstruction, atom maps, grain boundary character, and composition of the region close to the apex of the needle-shaped sample is shown in Figure 4. In the analyzed volume, a strong segregation tendency of Cu along general high angle grain boundaries (HAGBs) is observed, whereas Si atoms preferentially decorate low angle grain boundaries (LAGBs) and a high angle boundary with a misorientation of 20°. We mostly find low energy Σ3 boundaries in the volume analyzed here and the other interfaces are non-CSL type grain boundaries.

**Figure 4: Grain boundary character and segregation correlation at general boundaries**

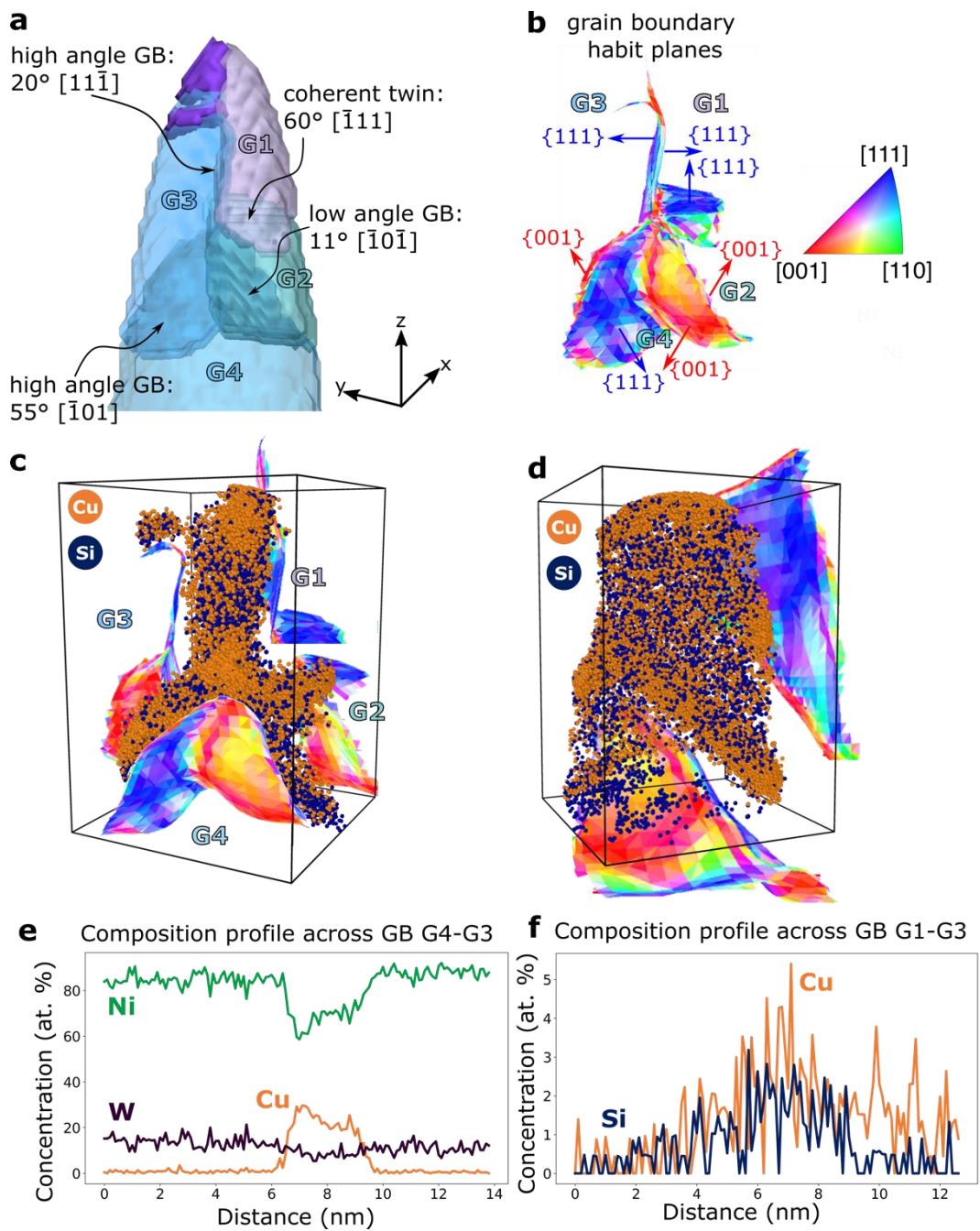

***Figure 4**: (a) 3D crystal orientation of the tip of the needle shaped sample with indicated grain boundary (GB) types, their corresponding misorientation angle, and global grain boundary habit plane. (b) Grain boundary habit plane (GB-HP) map colored according to the inverse*

*pole figure color code. The GB-HP map consists of two closely spaced surfaces corresponding to the GB plane normal of the adjacent crystals. (c) Exploded view of the GB-HP map showing the two surfaces superimposed on the corresponding 3D atom map of Cu and Si obtained by APT. (d) The same explosion-view as in (c) in a different view angle showing the GB between G1 and G3 plane on. (e) Concentration profile extracted across the high angle GB G3-G4 showing strong Cu enrichment. The Si profile is not shown here since it does not show any significant sign for segregation. (f) Concentration profile extracted across the high angle GB G1-G3 with a slight indication of Cu and Si segregation.*

A coherent $\Sigma 3$ 60° {111} twin boundary is observed between grains G1 and G2, which does not show any sign of solute segregation as expected[42]. The low angle GB surrounded by grains G2 and G4 with a misorientation angle of 11° exhibits similar GB plane normals of the abutting grains of $\{100\}_{G2}/\{100\}_{G4}$, which is primarily decorated by Si atoms and no Cu segregation is found (see supplementary Fig. S3 (a)). Regularly spaced peaks of Si are observed in the Si atom density map of this low angle GB, which suggests that Si atoms are segregating to the array of dislocations in the interface as shown in supplementary Fig. S3 (b). Strong Cu segregation of more than 30 at.% in peak concentration is observed at a general high angle GB with misorientation angle of ~55° in between grains G4 and G3 (GB plane normals $\{111\}_{G4}/\{100\}_{G3}$) as shown in Fig 4e. Interestingly, W seems to be slightly depleted at the GB and the Si concentration profile (see supplemental Fig. S4) does not show any significant sign of segregation. Such strong Cu segregation to a general high angle GB having nearly low-indexed GB planes, deviates from previous findings, where typically low segregation values are found for such types of interfaces[43,44]. The high angle GB separating grains G1 and G3 has a global GB plane normal close to $\{111\}_{G1}/\{111\}_{G3}$ and a misorientation angle of ~20°. This GB is decorated with both Cu and Si atoms as can be seen in the concentration profile shown in Fig 4f, but with a much lower content of the solutes compared to the GB between grains G3 and G4. The lower content of Cu at the 20° compared to the 55° GB could be related to several factors including the density of secondary GB dislocations, the GB habit plane normals and local GB parameters.

**Figure 5: Grain boundary character and segregation correlation at special boundaries**

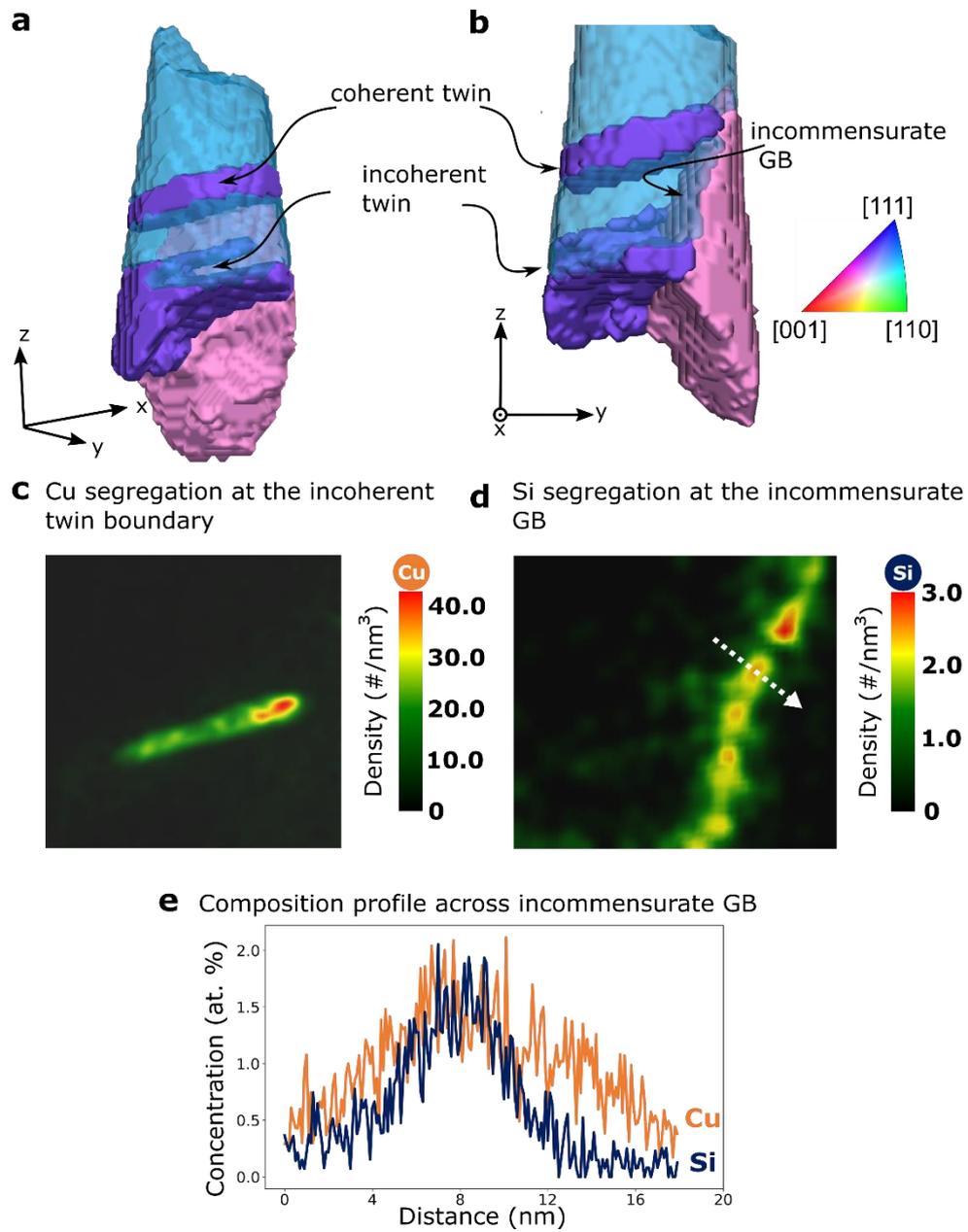

**Figure 5:** *(a) A cropped region from 3D reconstructed grains highlights the front (translucent) and back (solid) grains colored according to the inverse pole figure color code pointing along z-axis. (b) Grain boundaries of interest (coherent and incoherent twin boundaries, and incommensurate grain boundary) are marked. (c) and (d) shows Cu atoms and Si atoms density map respectively, on a 2D plane perpendicular to its iso-concentration surface placed such that the decoration of Cu and Si atoms can be visualized. In (d), a discontinuous enrichment up to 3 atoms/nm$^3$ of Si can be visualized. (e) 1D compositional profile measured across the incommensurate Gb in the direction of white arrow in (d) showing the maximum enrichment of Cu and Si at this boundary.*

Figures 5 a-e illustrate the 3D structure-composition correlation at a stepped twin boundary. The coherent twin boundary between the purple and blue grains shows no segregation, whereas a nanometer scale incoherent twin boundary segment is strongly enriched in Cu along its entire length (see Figure 5c). A GB between the purple and pink grains in Figure 5b is identified as an incommensurate (singular) boundary, a special asymmetric grain boundary with 45° misorientation and GB habit planes of (001)/(011), as discussed by Lejček et al.[45]. The special nature of this high angle GB seems to limit the segregation of both Cu (~1.5 at%) and Si (~ 1.5 at %) (see Figure 5e). A discrete modulated segregation pattern of Si atoms is observed (see Figure 5d), which can indicate a nanometer scale faceting of the boundary.

By correlating 3D crystal orientation mapping and APT on the same sample, we could elucidate the nature of the local GB structure and composition in a nanocrystalline material. The rich multidimensional data provides novel insights into the segregation behavior with nanometer resolution in three dimensions and provides access to a large fraction of GB-types, but is also capable to reveal often overlooked features, such as a nanometer sized incoherent GB segment. This experimental evidence lays the foundation to build atomistic and thermodynamic models to understand the underlying segregation mechanisms. For example, W segregation was only observed near triple junctions and to GBs in their close proximity. It seems that Cu displaces W during the annealing treatment from most of the GBs and also impacts the extent of Si segregation. Furthermore, the GB character dependent Cu segregation here deviates from the predicted GB segregation energy spectra of Cu at GBs in Ni polycrystals using the embedded atom potential method[46,47]. Such discrepancies between atomistic simulations and experimental observations require more holistic characterization of GBs and their related segregation behavior to ultimately build better models to predict interfacial properties. Our direct 3D correlation of GB structure and compositions opens this field and provides pathways to study multi-elemental segregation at complex 3D GB networks.

**Discussion**

In summary, a framework to probe the 3D nature of the structural parameters and composition of GBs is presented. We combine 4D-SPED tomography (3D crystal orientation mapping) with atom probe tomography (3D compositional information) to obtain a correlation of the macroscopic degrees of freedoms of interfaces along with their compositional fingerprint. In particular, we demonstrate a methodology to determine the 3D GB habit plane network on a quantitative basis, which is the foundation for understanding how the grain boundary character affects elemental segregation, which ultimately determines the properties of a material. A structurally and chemically complex nanocrystalline material is chosen to show the strength of the 3D correlative framework. Besides the local GB character, salient features such as incoherent twin boundary segments can be revealed both structurally and chemically by unleashing the strength of each of the techniques. The 3D structure-composition data opens the door to study complex interfacial networks in polycrystalline nanomaterials at nanometer resolution and provides novel insights into complex materials required for making accurate predictions of interfacial properties and with this material behavior. Another advantage is that the APT reconstruction can be calibrated using the 3D crystallographic information from 4D-

SPED tomography data. In many cases, crystallographic information in complex materials can hardly be discerned in APT due to the complex nature of the field evaporation process. In the future, it can be envisioned to use the 3D crystallographic data to locally refine the APT reconstruction on a 3D voxel basis by data fusion techniques. Advancements in electron detector technologies for 4D-STEM will allow to significantly reduce data acquisition times to eventually probe GB segregation phenomena during annealing treatments or electrochemical testing. Our results show the great potential of this new technique in the field of correlative microscopy and its application to grain boundary engineering to facilitate the development of advanced material systems.

**Methods**

**Materials**

An electrodeposited nanocrystalline (nc) Ni-14 at% W alloy on a Cu substrate was received from Xtalic Corporation, USA. The electrodeposition route produces a randomly oriented nano-grained material[48]. The sample was annealed at 600 °C for 6 h to crystallize the initially amorphous material in the as-deposited state and to initiate diffusion of Cu and Si along the grain boundaries of Ni-W nc-alloy.

**Correlative sample preparation using focused ion beam (FIB)**

Needle-shaped sample preparation for the correlative experiment was carried out using a dual-beam focused ion beam (FIB) instrument (Thermo Fisher Scientific Helios Nanolab 600i). The sample was lifted out and mounted onto the APT-compatible cylindrical Cu post of the Fischione on-axis rotation tomography holder. The apex of the Cu post was pre-sharpened using FIB to mount the sample. Subsequently, the sample was thinned down to a needle shape at 30 kV accelerating voltage using a Ga ion beam with a current range from 2.5-0.8 nA. A final cleaning procedure at 2 kV and 16 pA current was carried out to remove severely damaged regions during thinning at high energy (30 kV).

**4D-scanning precession electron diffraction tomography**

The 4D-SPED was performed in a JEOL JEM-2200FS microscope operated at an accelerating voltage of 200 kV using a 10 µm condenser (CL1) aperture and a calibrated camera length of 400 mm (calibrated using ASTAR) using 30 eV energy filter. The probe size was estimated to be ~2.5-3 nm after a precession angle of 1° at 100 Hz was applied by the Digistar hardware unit (NanoMEGAS SPRL). The precessed nano-probe beam is scanned across the needle-shaped sample with a step size of 2.4 nm and nanobeam diffraction patterns are acquired at each probe position with a 4k x 4k CMOS detector (TemCam-XF416-TVIPS) hardware-binned to 512 x 512 pixels². The diffraction patterns were further binned numerically to 256 x 256 pixels² using the tvipsconverter software for further processing[49]. To speed-up the acquisition time, the needle shaped sample was aligned along the fast scan direction and a frame comprising 100 x 276 pixels² was used to scan the sample. A pixel dwell time of 41 ms was used to capture each nanobeam diffraction pattern, resulting in ~4 rotations of the precessed incident beam per frame, which significantly reduces dynamical diffraction effects. A 4D-SPED scan was recorded over a sample tilt range from -80° to +80° with a tilt step increment of 10°, for a series of 17 projections in total. A Fischione Instrument Model 2050 On-Axis Rotation Tomography Holder was used to acquire the tilt series. This results in a complex 3D 4D-SPED dataset consisting of 469,200 precessed, nanobeam electron diffraction patterns in total.

**Atom probe tomography**

After the 4D-SPED tomography acquisition, the sample was showered by low-kV Ar ions in a PECS Model 682 system by Gatan at 2 kV and 32 µA current to remove hydrocarbon layers formed during the TEM measurement before loading the sample into the APT analysis chamber[50]. The APT measurement was conducted using an atom probe 5000XR instrument (LEAP, Cameca Instruments). Laser pulsing mode was applied at a pulse repetition rate of 200 kHz and a pulse energy of 40 pJ. The sample's base temperature was kept at 70 K and a target detection rate of 1% was set. Data analysis was performed using the APsuits software package.

**Data Processing and Analysis**

The 4D-SPED data was processed using the ASTAR software package and in-house developed Python codes mainly relying on the NumPy, SciPy, and orix software packages[51–53]. The azimuthal rotation of the diffraction pattern with respect to the rotation axis of the sample holder was calibrated using the procedure described in the paper[16]. The computed rotation vector (axis and angle) couples the tilt-series. A multi-indexing approach was employed to retrieve through-thickness information and to treat overlapping grains[34]. Orientation-specific virtual apertures were selected for generating the projected virtual dark field images of individual grains throughout the tilt series. A coarse alignment of the projection images in the tilt series is performed manually in the Tomviz software[54]. Fiducial alignment as implemented in the IMOD software was employed for fine alignment before tomographic reconstruction using the "Simultaneous Iterative Reconstruction Technique" (SIRT) algorithm. Tomviz and Paraview were used for 3-dimensional visualization and rendering[55]. In order to perform correlative data analysis, the axes were defined according to APT convention.

**Acknowledgements**

S.M.D., M.H., and X.Z. acknowledge funding by the German Research Foundation (DFG) via project HE 7225/11–1. X.Z. is supported by the Alexander von Humboldt Stiftung. P.H., W.L, and E.F.R. would like to recognize funding from the Agence Nationale de la Recherche (grant no. ANR-19-CE42–0017).

# Supplementary Information

# Correlating grain boundary character and composition in 3-dimensions using 4D-scanning precession electron diffraction and atom probe tomography

Saurabh M. Das[1*+], Patrick Harrison[2+], Srikakulapu Kiranbabu[1], Xuyang Zhou[1], Wolfgang Ludwig[3,4*], Edgar F. Rauch[2], Michael Herbig[1], Christian H. Liebscher[1,5,6*]

[1] Max-Planck-Institut for Sustainable Materials (Max-Planck-Institut für Eisenforschung), Max-Planck-Straβe 1, 40237 Düsseldorf, Germany
[2] SIMAP Laboratory, CNRS-Grenoble INP, BP 46 101 rue de la Physique, 38402 Saint Martin d'Hères, France
[3] ESRF–The European Synchrotron, 71 Av. des Martyrs, 38000 Grenoble, France
[4] MATEIS, INSA Lyon, UMR 5510 CNRS, 25 av Jean Capelle, 69621 Villeurbanne, France
[5] Research Center Future Energy Materials and Systems, Ruhr Univeristy Bochum, Universitätsstr. 150, 44801 Bochum, Germany
[6] Faculty of Physics and Astronomy, Ruhr Univeristy Bochum, Universitätsstr. 150, 44801 Bochum, Germany

*Corresponding authors,

email: 100rabh2992@gmail.com (Saurabh Mohan Das), wolfgang.ludwig@esrf.fr (Wolfgang Ludwig),

christian.liebscher@rub.de (Christian H. Liebscher)

[+]These authors contributed to the work equally.

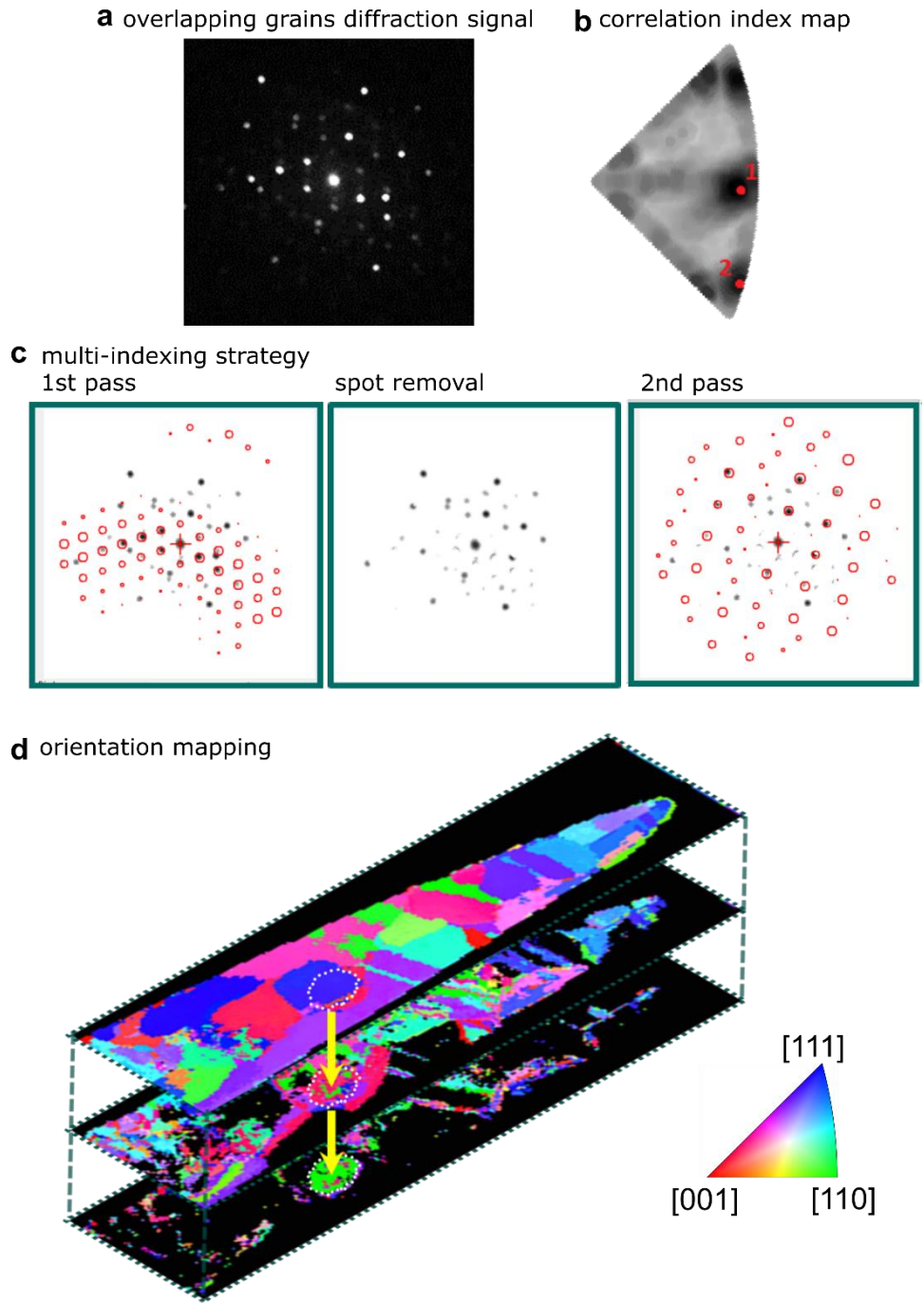

**Figure S1:** Multi-indexing strategy reveals the hidden grain along the beam direction. (a) A diffraction pattern from the 60° tilt dataset containing signal from overlapping grains. (b) Correlation index map showing the two possible solutions marked as 1 and 2. (c) ACOM best-matched template overlay (left), the new diffraction pattern after removing the diffraction signal corresponding to the best-matched template (middle), and (right) the 2nd best-matched template corresponds to the non-dominant grain. (d) Orientation mapping of a 60° tilt data set taken as an example of multi-indexing strategy reveals a grain

(highlighted with a dotted white circle) with orientation close to (110) that only is indexed in the 2$^{nd}$ and 3$^{rd}$ indexing passes.

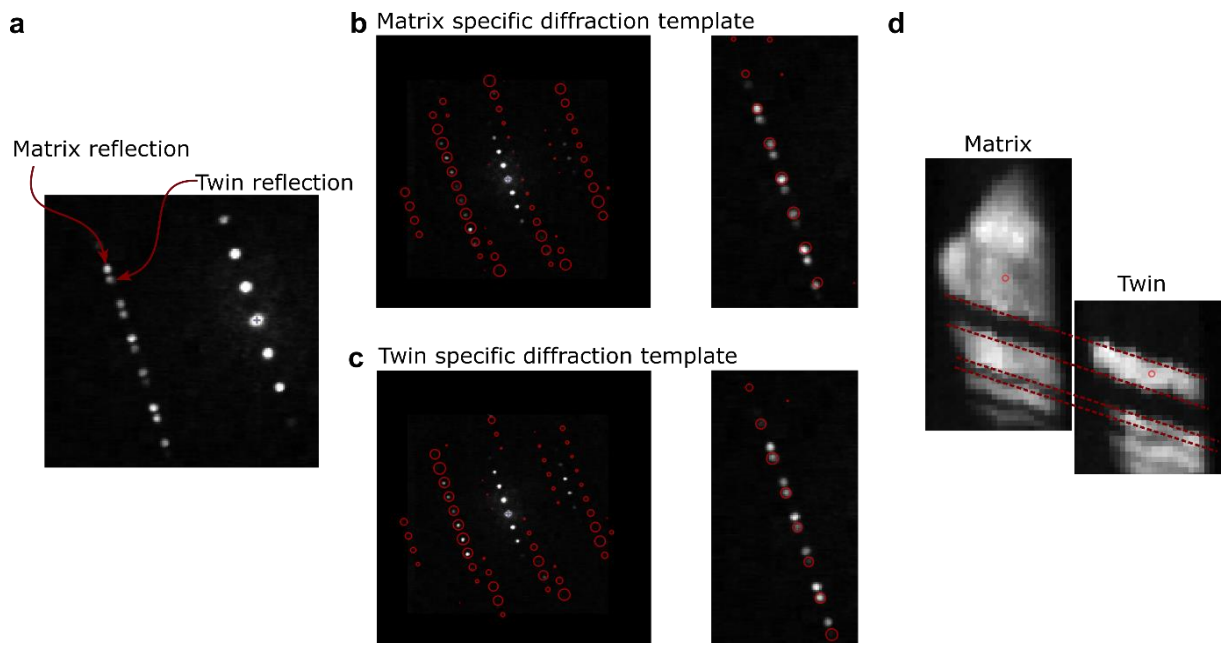

**Figure S2:** Twin reconstruction strategy. (a) Cropped region of diffraction pattern showing the matrix and twin reflections. (b) Matrix-specific diffraction template, and (c) twin-specific diffraction templates are used to separate the matrix and twin reflections, and (d) subsequently these specific templates are used as virtual apertures to generate the matrix and twin projection images using Virtual Dark Field imaging algorithm[1].

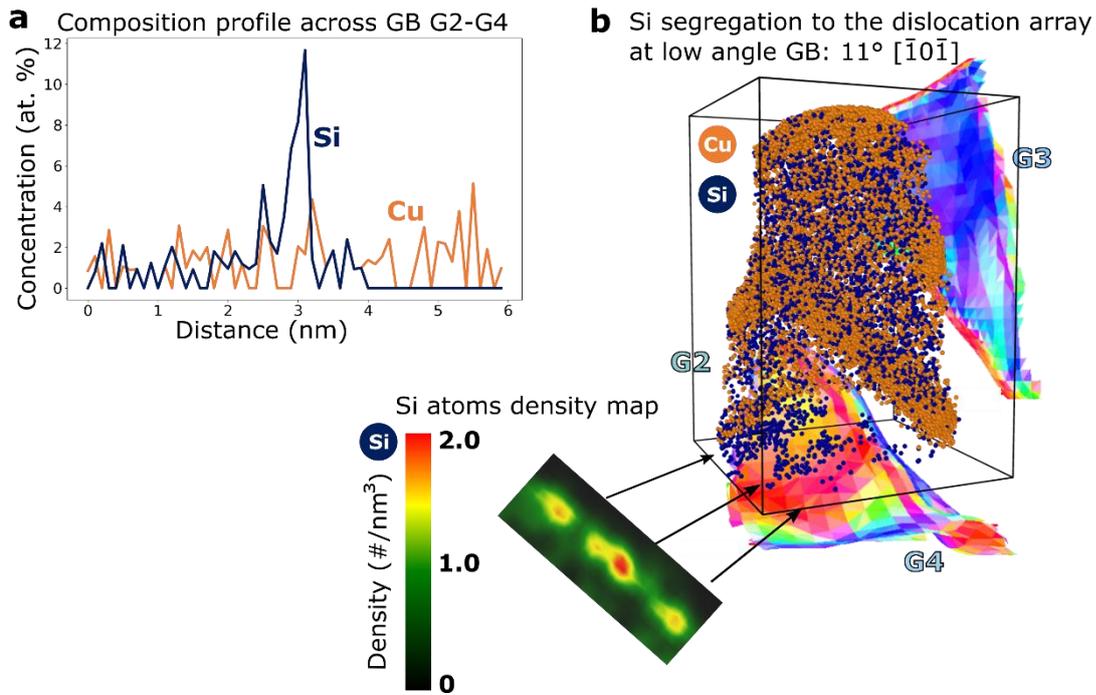

**Figure S3:** Si atoms are segregating to the array of dislocations at G2-G4 low angle GB. (a) Si and Cu concentration profiles across G2-G4 GB show only Si enrichment at the dislocation. (b) Si atoms density map on a 2D plane perpendicular to its iso-concentration surface across G2-G4 GB, placed such that the decoration of dislocation core is visualized.

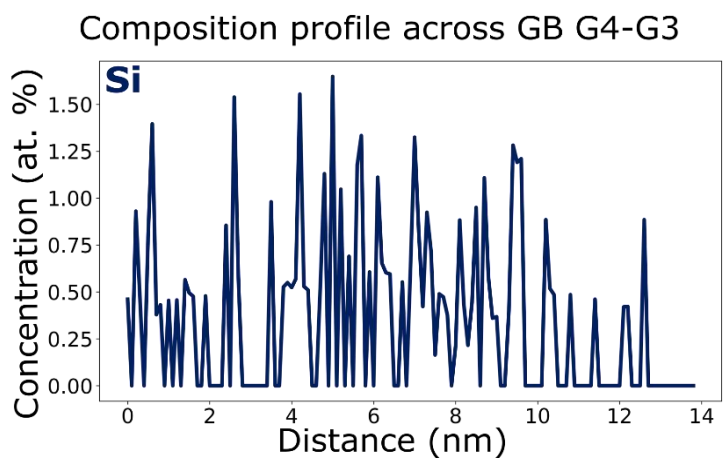

**Figure S4:** Si concentration profile does not show any significant sign of segregation across G4-G3 high angle GB.